# Proposal for Adding Useful Features to Petri-Net Model Checkers

Hubert Garavel

*Univ. Grenoble Alpes, INRIA, CNRS, Grenoble INP, LIG, 38000 Grenoble, France*

version 12 - December 23, 2020

## Abstract

*Solutions proposed for the longstanding problem of automatic decomposition of Petri nets into concurrent processes, as well as methods developed in Grenoble for the automatic conversion of safe Petri nets to NUPNs (Nested-Unit Petri Nets), require certain properties to be computed on Petri nets. We notice that, although these properties are theoretically interesting and practically useful, they are not currently implemented in mainstream Petri net tools. Taking into account such properties would open fruitful research directions for tool developers, and new perspectives for the Model Checking Contest as well.*

## 1. The Dead Places Problem

Definition: A place $p$ is *dead* iff there exists no reachable marking $M$ such that $p$ is marked in $M$.

Definition: Given a Petri net, the *Dead Places Problem* consists in finding all the dead places. Equivalently, if P is the set of places, this problem consists in computing a vector of |P| bits, each of which is 1 iff the corresponding place is dead.

This problem is the equivalent, for Petri nets, of the dead code problem in software engineering. This is a practically relevant problem: dead code is a nuisance for maintenance, and most software methodologies require to get rid of dead code. Moreover, many global properties of a network can be changed to true or to false just by adding dead places, so it is important to detect and eliminate such places to only consider a truly-minimal Petri net. For instance, in industrial automation, the Grafcet specification prohibits Sequential Function Charts containing "unreachable" branches (i.e., Petri nets with dead places or transitions).

In model-checking verification, dead places are likely to increase the memory cost of verification. Moreover, even if nets contain only a few dead places in practice, these may cause trouble when performing structural analyses or transformations of Petri nets, as they may invalidate certain "good" properties (e.g., free choice) and thus lead to incorrect transformations, or prevent the application of efficient algorithms relying on such properties.

## 2. The Dead Transitions Problem

Definition: A transition $t$ is *dead* iff there exists no reachable marking $M$ such that $t$ is enabled in $M$.

Definition: Given a Petri net, the *Dead Transitions Problem* consists in finding all the dead transitions. Equivalently, if T is the set of transitions, this problem consists in computing a vector of |T| bits, each of which is 1 iff the corresponding transition is dead.

This problem is relevant for the same reasons as for the Dead Places Problem.

It is different from another problem : "is the Petri net quasi-live?", since that latter problem only requires a Boolean answer, whereas the Dead Transitions Problem asks for a vector of Booleans. In fact, the Dead Transitions Problem is more general, as the network is quasi-live iff the answer to the Dead Transitions Problem is the vector of zeros. In practice, when debugging a complex network, the quasi-liveness property does not suffice: knowing the existence of dead transitions is not enough, one needs to have the list of dead transitions.

## 3. The Concurrent Places Problem

Definition: Two places *p* and *p'* are concurrent iff there exists a reachable marking *M* such that both p and p' are marked in *M*.

Definition: Given a Petri net, the *Concurrent Places Problem* consists in finding all pairs of concurrent places. Equivalently, if P is the set of places, given that the concurrency relation is symmetric, this problem consists in computing a half matrix of |P|.(|P|+1)/2 bits, each of which is 1 iff the corresponding pair of places are concurrent.

This problem generalizes the Dead Places Problem. Indeed, a place is dead iff it is not concurrent with itself.

This problem is practically relevant, since the concurrency relation characterizes those net parts that can be simultaneously active. It is mentioned in many publications under various names, such as: *coexistency defined by markings* [Jan84, Sect. 9], *concurrency graph* [Kar12] [WKAK14], or *concurrency relation* [Kov92] [SY95] [KE96] [Kov00] [GS06], etc. These definitions slightly differ by minor details, such as the kind of Petri nets considered, or the handling of reflexivity, i.e., whether and when a place is concurrent with itself or not.

## 4. Comparison with temporal logics

Interestingly, the three above problems can be expressed using temporal logics (e.g., CTL or LTL) as reachability formulas following the same pattern. Given a set of places *M*, let *R* (*M*) be the predicate: "does it exist a reachable marking containing all places of *M*?".

- Dead Places Problem: for each place *p*, compute $R(\{p\})$.
- Dead Transitions Problem: for each transition *t*, compute $R(\bullet t)$.
- Concurrent Places Problem: for each pair (*p*, *p'*) of places, compute $R(\{p, p'\})$.

It would be therefore tempting to express these problems as particular cases of the evaluation of CTL or LTL formulas. However, we believe that such reduction is not the best option or, at least, it is not the only option:

- For tool users, it is much easier to directly invoke a Petri-net tool that has built-in options (such as **-dead-places**, -**dead-transitions**, or **-concurrent-places**) rather that building a set of temporal-logic formulas, invoking a model checker on each of these formulas, then collecting and aggregating the results of all these invocations in one single file — keeping in mind that the number of these formulas is linear or even quadratic in the size of the Petri net, thus leading to a huge input file or to a large number of input files.

- For tool developers who plan to solve these three problems by means of LTL or CTL model checking, it is more efficient to have built-in options (such as **-dead-places**, etc.) and internally generate and evaluate the set of temporal-logic formulas: so doing, the tool avoids parsing costs, as the formulas can be directly generated in abstract syntax rather than concrete syntax; moreover, the tool knows that all these formulas are correct by construction and, thus, can avoid checking their correctness, which would not be possible if these formulas were provided by a human user.

- Also, the existence of built-in options opens the way to strategies that would not be fruitful when evaluating temporal-logic formulas one by one. If a tool knows it has to evaluate thousands of formulas on the same model, then it can profitably try applying preliminary simplifications (e.g., structural transformations) or sophisticated analyses to this model. Similarly, the tool may profitably consider using *global model checking* (i.e., explicitly building the reachable state space first, then evaluating all formulas on this state space) whereas, for a single formula, local model checking (i.e., on the fly evaluation) might be preferred. One may also consider combined aproaches, e.g., first performing global model checking to get as many results as possible, then spotting the missing results and using local model checking to get them one by one, or two by two, etc.

Thus, although the three above problems can be reduced to the evaluation of a set of temporal-logic formulas, it is better to express these problems at a higher level (namely, equip Petri-net tools with built-in options for these problems) and let tool developers choose which approaches solve these problems best. Said otherwise, temporal logic is a possible way to express and solve these problems, but it should not be the mandatory way.

## 5. An example: Implementation in the CAESAR.BDD tool

CAESAR.BDD [Cae19] is a Petri-net verification tool that is part of the CADP toolbox [Cad19]. The recent versions of CAESAR.BDD provide three options **-dead-places**, **-dead-transitions**, and **-concurrent-places** that address the three aforementioned problems. In practice, the **-dead-transitions** option is used to simplify large interpreted Petri nets automatically generated from specifications written using higher-level languages such as LOTOS, LNT, AADL, etc. The **-concurrent-places** option is used to automatically convert ordinary, safe Petri nets into NUPNs (*Nested-Unit Petri Nets*) [Gar19] by inferring concurrency, locality, and hierarchy information from any given net [BGP20]. CAESAR.BDD does not support temporal-logic formulas, but provides many other features, among which a **-mcc** option that automatically generates LaTeX model forms for the Model Checking Contest.

CAESAR.BDD. explores the reachable state space of a Petri net using the BDD package CUDD [Cudd19]. Once the state space has been generated, the results of **-dead-places, -dead-transitions** and **-concurrent-places** are computed by checking the predicate *R* defined in Section 4. If the state space has been incompletely generated, "*unknown*" results are returned for certain places, transitions, or pairs of places. Additional algorithms are implemented to reduce the proportion of "*unknown*" results.

The **-dead-transitions** and **-concurrent-places** options of CAESAR.BDD have been assessed in 2019 on a large base of 13556 Petri nets or NUPNs from diverse origins, including many nets taken from the model collection of the Model Checking Contest. These experiments have been conducted on Grid 5000, a French national grid. More than 530 years of CPU have been used (including other experiments, failed runs due to mistakes, runs halted on timeout, etc.). CAESAR.BDD was able to solve the Dead Transitions Problem for 13225 models (97.6%) and to solve the Concurrent Places Problem completely for 13176 models (97.2%); on 363 of the remaining models (2.68%), CAESAR.BDD could produce approximate results (true, false, or unknown) for the Concurrent Places Problem.

## 6. Potential interest for the Model Checking Contest

It would be interesting to know whether other tools, possibly implementing different techniques, could perform better than CAESAR.BDD on these challenging problems, either by tackling models that could not be handled by CAESAR.BDD, or by computing solutions with less "*unknown*" results, or by providing equivalent results in shorter time with a lower computation cost.

Should enough Petri-net tools implement algorithms for the three above problems (which should be easy, at least for the tools supporting temporal-logic formulas), then the Model Checking Contest (MCC) [Mcc19] could include these problems — most likely in its "Global Properties" competition category, which has been recently introduced and is currently limited to the "Deadlock" problem. So doing, the MCC would foster research on new problems and allow a fair comparison of the tool performances on these problems.

At present, all the tools regularly participating in the MCC have converged to give nearly 100% of correct answers. Time has come to propose new challenges. For the MCC itself, the advantages would be multiple:

- **Closer to user needs:** The current situation, where tools have to evaluate only 16 formulas for each problem, is merely an approximation of the real needs of the user. Considering, for instance, the Dead Transitions Problem, a tool user will not be satisfied by knowing that a net is not quasi-live, nor by knowing whether 16 transitions taken randomly in this net are quasi-live or not; the user expects the tool to provide the complete list of dead transitions (or, at least, as much as possible dead transitions within a given lapse of CPU time).

- **Beyond temporal logics:** the MCC is currently very much oriented towards temporal logics (actually, state-based logics, as action-based logics with transition-oriented properties are not supported). The proposed challenges, which would be expressed as global problems without explicit reference to temporal logics, might attract new tools that, like CAESAR.BDD, perform state-space exploration to answer useful queries not expressed in temporal logic.

- **Beyond local model checking:** the MCC currently favors local (i.e., on-the-fly) model checking, where only the fragment of the state space relevant to evaluate a given property is evaluated. Global model checking is implicitly discouraged, as exploring the state space entirely may be too expensive if only 16 formulas are to be evaluated on this state space. However, global model checking may be of interest when evaluating a larger number of formulas: for instance, global model checking has been recently used during the RERS 2019 contest to evaluate 360 temporal-logic formulas in about 10 hours using the CADP toolbox [Rers19]. The proposed evolution of the MCC would address this issue by proposing a better balance between local and global model checking, the former no longer being the default approach.

- **Enabling structural reductions:** for the same reasons, the proposed evolution of the MCC would provide tool developers with greater incentive to perform structural reductions, so as to simplify the state space, still preserving its key properties, before evaluating temporal logic formulas. Currently, some tools prefer restarting from scratch the state-space exploration everytime a new property is evaluated. Such a strategy is only possible if the number of temporal logic formulas for the same model is low.

Technically, the introduction of the three aforementioned problems would not change the MCC rules. As for the other competition categories, each tool would produce a vector of values (False, True, Unknown, Crash, Timeout), the only difference being that the size of this vector is not constant (i.e., 16 when 16 formulas are evaluated) but depends on the size of the model itself (e.g., linear in the number of transitions or quadratic in the number of places).

## 7. Proposed output format for the Dead Places and Dead Transitions Problems

In order to allow a simple comparison of the outputs of the various tools, we suggest to adopt a precisely-defined file format for the outputs of the Dead Places and Dead Transitions Problems.

This format should be kept as simple as possible, avoiding XML-like syntax, which is likely to cause file-size explosion in the case of the Concurrent Places Problem (see Section 8 below).

For the Dead Places Problem, all the places of the PNML file should be sorted in unambiguous order and numbered from 1 to |P|, e.g., according to their order of declaration in the input PNML file. Then, the output of the Dead Places Problem should be a text file containing one line of |P| characters. The *i*-th character should be '1' if the *i*-th place is dead, '0' if it is not dead, or '.' if this information is unknown because the state space was incompletely explored (e.g, halted due to a timeout).

Note 1: We prefer using '.' rather than '?' since the latter character has the same height as '0' and '1', and was found to be visually painful and error-prone for human readers.

Note 2: Initially, each of the |P| characters was on a separate line, because some editors (e.g., vi) and text processors (such as awk, grep, or sed) have fixed-size buffers and may face problems handling long lines. Yet, putting all characters on one single line was later preferred, since it provides human readers with a global overview of the output contents.

Finally, the output line of |P| characters is compressed using the simple algorithm based on run-length encoding: if a given character *c* is repeated *n* times, with *n* > 3, the *n* consecutive occurrences of this character *c* are replaced with the abbreviated notation *c(n)*. This algorithm is detailed in the section entitled "COMPRESSION SECTION" of the following Web page: http://cadp.inria.fr/man/caesar.bdd.html. In practice, this compression achieves significant savings for large Petri nets having thousands of places, only a few of which are dead.

For the Dead Transition Problem, we propose the same format, replacing "place" by "transition" and |P| by |T|.

## 8. Proposed output format for the Concurrent Places Problem

Given that the concurrency relation between places is symmetric, it is sufficient to represent this relation as a lower half of a matrix, whose lines and columns are indexed by the unique numbers assigned to each place (see Section 7). This half matrix has |P|.(|P|+1)/2 cells. Each cell of the matrix contains a single character: '1' if the corresponding places are concurrent, '0' if they are not, or '.' if the information is unknown.

Note: The **-concurrent-places** option currently implemented in CAESAR.BDD is slightly different, as matrix cells may contain other character values, e.g., '=', '<', and '>', which mean "non-concurrent" with the extra information that the two places are located in the same NUPN unit or in parent NUPN units.

Experiments conducted with CAESAR.BDD on large Petri nets taken from the MCC collection have shown that the half matrix can become quite large (files of several gigabytes). For this reason, each output line of the half matrix is compressed using the aforementioned algorithm described in the section entitled "COMPRESSION SECTION" of the Web page: http://cadp.inria.fr/man/caesar.bdd.html. The table given in Annex A shows an uncompressed matrix (left) and its compressed version (right). Measured on 12671 models, the average compression factor was 214; on large models, the compression factor reached higher values; for instance, a 8.6-Gbyte file was compressed to a 2-Mbyte file (compression factor: 4270).

## Acknowlegements

## References

[BGP20] Pierre Bouvier, Hubert Garavel, and Hernan Ponce de Leon. *Automatic Decomposition of Petri Nets into Automata Networks - A Synthetic Account*. In Ryszard Janicki, Natalia Sidorova, and Thomas Chatain, eds., Proceedings of the 41st International Conference on Application and Theory of Petri Nets and Concurrency (PETRI NETS'20), Paris, France. Lecture Notes in Computer Science, vol. 12152, pages 3-23, Springer, June 2020.

[Cad19] http://cadp.inria.fr

[Cae19] http://cadp.inria.fr/man/caesar.bdd.html

[Cudd19] http://vlsi.colorado.edu/~fabio/CUDD

[Gar19] Hubert Garavel. *Nested-Unit Petri Nets*. Journal of Logical and Algebraic Methods in Programming, vol. 104, pp. 60–99, 2019. http://cadp.inria.fr/publications/Garavel-19.html

[GS06] Hubert Garavel and Wendelin Serwe. *State Space Reduction for Process Algebra Specifications*. Theoretical Computer Science, vol. 351(2), pp. 131-145, February 2006.

[Jan84] Ryszard Janicki. *Nets, Sequential Components and Concurrency Relations*. Theoretical Computer Science, vol. 29, pp. 87-121, 1984.

[Kar12] Andrei Karatkevich. *Conditions of SM-Coverability of Petri Nets.* September 2012.
https://www.researchgate.net/publication/267508814_Conditions_of_SM-Coverability_of_Petri_Nets

[KE96] Andrei Kovalyov and Javier Esparza. *A Polynomial Algorithm to Compute the Concurrency Relation of Free-Choice Signal Transition Graphs*. Proceedings of the 3rd Workshop on Discrete Event Systems (WODES'96), Edinburgh, Scotland, UK, August 1996.

[Kov92] Andrei V. Kovalyov. *Concurrency Relations and the Safety Problem for Petri Nets*. Proceedings of the 13th International Conference on Application and Theory of Petri Nets, Lecture Notes in Computer Science, vol. 616, 1992.

[Kov00] Andrei V. Kovalyov. *A Polynomial Algorithm to Compute the Concurrency Relation of a Regular STG.* In A. Yakovlev et al. (eds.), *Hardware Design and Petri Nets*, chapter 6, pp. 107-126, Kluwer Academic Publishers, 2000.

[Mcc19] http://mcc.lip6.fr

[Rers19] http://cadp.inria.fr/news12.html

[SY95] Alexei Semenov and Alexandre Yakovlev. *Combining Partial Orders and Symbolic Traversal for Efficient Verification of Asynchronous Circuits*. In Tatsuo Ohtsuki and Steven Johnson, eds., Proceedings of the 12th International Conference on Computer Hardware Description Languages and their Applications (CHDL'95), Makuhari, Chiba, Japan, IEEE, August-September 1995.

[WKAK14] Remigiusz Wisniewski, Andrei Karatkevich, Marian Adamski, and Daniel Kur. *Application of Comparability Graphs in Decomposition of Petri Nets*. Proceedings of the 7th International Conference on Human System Interactions (HSI'14), Costa da Caparica, Portugal, pp. 216-220. IEEE, June 2014.

## Appendix A

The table below gives the half matrix for the Concurrent Places Problem. The uncompressed matrix is shown on the left and its compressed version on the right.

```
1                                    1
11                                   11
111                                  111
0111                                 0111
10011                                10011
100111                               100111
0111111                              01(6)
00011001                             00011001
100111111                            1001(6)
1001111111                           1001(7)
10011110111                          1001(4)0111
100111110011                         1001(5)0011
1001111100111                        1001(5)00111
10011110111111                       1001(4)01(6)
100111100011001                      1001(4)00011001
0000000011111111                     0(8)1(8)
00000000111111111                    0(8)1(9)
100111101111110111                   1001(4)01(6)0111
1001111011111110011                  1001(4)01(7)0011
10011111111111100111                 1001(12)00111
100111000011011111111                1001(4)0(5)1101(6)
000000000110000000011                0(10)110(8)11
0000000000110000000111               0(10)110(8)111
100111100111101111111001             1001(5)001(4)01(6)001
1001110111111111111110011            1001(4)01(13)0011
10011101111111111110111111           1001(4)01(12)01(6)
100111111111101111111001111          1001(11)01(6)001(4)
1001111001111011111011111111         1001(5)001(4)01(5)01(7)
10011110011110111110111111111        1001(5)001(4)01(5)01(8)
100111000011111111000100111111       1001(4)0(5)1(8)0001001(4)
1001110000011111111000100111111      1001(4)0(5)1(8)0001001(5)
00000111111111000110100111011111     0(5)1(9)000110100111011(5)
000001111111110001101001110111111    0(5)1(9)000110100111011(6)
01111110111111001110011111111111     01(6)01(7)001(4)001(11)
011111111111110011100111111110011    01(14)001(4)001(8)0011
0111111111111100111001111111100111   01(14)001(4)001(8)00111
1001111001111111110111111110011111   1001(5)001(10)01(8)001(6)
10011101111111111111001111100111111  1001(4)01(13)001(6)001(7)
100111011111011111100111111111111111 1001(4)01(6)01(6)001(16)
1001110111110111110011111111111111111 1001(4)01(6)01(6)001(17)
10011111111110111111001111111100111111111 1001(11)01(6)001(8)001(8)
```